\documentclass[10pt,a4wide]{article}

\usepackage[english]{babel}
\usepackage[sorting=none]{biblatex}
\usepackage[letterpaper,top=2cm,bottom=2cm,left=3cm,right=3cm,marginparwidth=1.75cm]{geometry}
\usepackage{float}
\usepackage{amsmath}
\usepackage{graphicx}
\usepackage{caption}
\usepackage{subcaption}
\usepackage{hyperref}
\hypersetup{colorlinks,allcolors=blue}
\usepackage{amsthm,amssymb,mathrsfs,setspace,booktabs}
\usepackage{mathtools,amsmath,nccmath}
\usepackage{authblk}
\usepackage{blindtext}
\DeclareUnicodeCharacter{202F}{\thinspace}
\begin{document}
\title{Loop-induced Higgs decays in two Higgs doublet scenarios: some observations}

\author[1]{Snehit Panghal}
\author[2]{Biswarup Mukhopadhyaya}
\affil[1]{Department of Physics, Indian Institute of Technology Bombay, Mumbai, Maharashtra 400076, India}
\affil[2]{Department of Physical Sciences, Indian Institute of Science Education and Research Kolkata,
Mohanpur, 741246, India}
\affil[1]{\href{mailto:23d1052@iitb.ac.in}{23d1052@iitb.ac.in}}
\affil[2]{\href{mailto:biswarup@iiserkol.ac.in}{biswarup@iiserkol.ac.in}}
\maketitle
\begin{abstract}
We study the decay of three neutral Higgs bosons into $\gamma\gamma$ and $Z\gamma$ in the four major types of two Higgs doublet models with CP conserved. Particular focus is made on the possibility of the width of the latter decay dominating over the former. For the 125-GeV scalar, such a possibility is mostly excluded when one imposes all theoretical and experimental constraints, other than the Higgs data at the Large Hadron Collider. For the other CP-even neutral scalar, there remain regions in each kind of model where the $Z\gamma$ mode dominates. For the CP-odd state, this possibility is restricted to only the models of Types II and Y, in certain regions of the parameter space.
\end{abstract}

\section{Introduction}
Ever since the discovery of a 125-GeV scalar at the Large Hadron Collider (LHC)\cite{ATLAS:2012yve,CMS:2012qbp}, we have accumulated enough data to be convinced that
this scalar plays a significant role in executing spontaneous breakdown of electroweak symmetry, as laid down in the standard model (SM) of particle physics\cite{Higgs:1964pj,PhysRevLett.13.321,PhysRevLett.13.585}. Whether the role is {\em exclusive} is, however, not completely clear yet. In principle, there may exist additional scalars belonging to doublets (or even higher representations) of SU(2), which participate in electroweak symmetry breaking (EWSB) following the same paradigm. While the issue could be largely settled by measuring precisely the quartic self-interaction strength of the 125-GeV scalar, such measurement is not an immediate possibility yet\cite{Agrawal:2019bpm}. It is, therefore, hardly surprising that additional scalar multiplets should be postulated and examined in the light of the available data. The minimal templates of such scalar sectors beyond the standard model (BSM) are two Higgs doublet models (2HDM)\cite{Branco:2011iw}.

A crucial test for any 2HDM scenario lies in the prediction of loop-induced decays of scalars, especially those of the 125-GeV particle, which we denote by $h$. Such decays which are not allowed at the tree-level are $h \rightarrow \gamma \gamma$ and $h \rightarrow Z \gamma$. While the former has already been measured to have a branching ratio (BR) of $(2.27 \pm 2.1\%) \times 10^{-3} $, the latter is decisively found to have smaller rates\cite{10.1093/ptep/ptac097,CMS:2023mku}, thus confirming that the diphoton channel has larger decay width. In this paper, we shall scan the parameter space 
of each of four major types of 2HDM, studying the BR for the diphoton mode vis-à-vis the $Z\gamma$ mode for not only $h$ but also the other neutral spin-0 field in each scenario.

The four types of 2HDM considered here are those of Type-I, Type-II,
Type-X and Type-Y (the flipped) kind\footnote{There are some other scenarios, in whose context the relative strength of $Z\gamma$ and $\gamma\gamma$ decay channels have been explored. See, for example, \cite{Benbrik:2022bol}.}. There are various motivations for such extensions, ranging from supersymmetry\cite{Martin:1997ns} to the anomalous magnetic moment of the muon\cite{Abe:2015oca,Kim:2022xuo}.The main difference among them lies in the nature of their Yukawa couplings to quarks and leptons. In each of them, there exist in addition to $h$, an additional neutral scalar $H$, a neutral pseudoscalar $A$, and a pair of mutually conjugate charged scalars $H^\pm$. While the additional scalars may be beyond current experimental reach, there are contributions of the charged
scalars even in $h \rightarrow \gamma\gamma, Z\gamma$. Thus, the 2HDM parameter space in each case plays a role in deciding the BR of each of these decays, as well as their relative strength. Moreover, the 2HDM parameter space is additionally limited by theoretical and experimental constraints. While the former category comprises  vacuum stability as well as perturbative unitarity, the latter includes direct search limits on the additional scalars, electroweak constraints (EWPC), and also those from rare decays in flavor physics, such as $b \rightarrow s\gamma$ and $B_s \rightarrow \mu^+ \mu^-$. And, last but not the least, the existing data on the particle $h$ imposes crucial on any BSM scenario including 2HDM.

The detailed study reported here takes all the above constraints into account. There emerges, in particular, a rather striking observation: {\em the diphoton channel for $h$ is  automatically predicted over most of the parameter space to have larger BR than $Z\gamma$ in all four 2HDM types, once all constraints  other than those from $h$-data are taken into account}. One thus concludes that the thus allowed regions for the four types of 2HDM are consistent with the observations on loop induced decays of the 125-GeV scalar. In addition, we also identify regions where $Z\gamma$ can dominate over diphotons in the decays of $H$ and $A$. This can help in identifying to which category of 2HDM such scalars belong, if they are revealed in the upcoming data.   

\section{The major types of 2HDM}
The 2HDM is the simplest extension of the SM Higgs, wherein there is an additional scalar $SU(2)_L$ doublet. The doublets can be expanded about their vacuum expectation values(vev) as\cite{Branco:2011iw,Bhattacharyya:2015nca,Haber:2015pua,Wang:2022yhm}
\begin{equation}
    \Phi_{1}=\left(\begin{array}{c}
\Phi_{1}^{+} \\
\frac{v_{1}+\rho_{1}+i \eta_{1}}{\sqrt{2}}
\end{array}\right) \quad , \quad \Phi_{2}=\left(\begin{array}{c}
\Phi_{2}^{+} \\
\frac{v_{2}+\rho_{2}+i \eta_{2}}{\sqrt{2}}
\end{array}\right), 
\end{equation}
where SM vacuum expectation value, $v=246$ GeV is expressed as $v = \sqrt{v_1^2 + v_2^2}$. Thus, unlike SM, there are eight real scalar fields in the model, rather than four. From Goldstone's theorem, three of the fields gives mass to the W and Z bosons because of the spontaneous symmetry breaking\cite{PhysRev.117.648}, and the rest of the five correspond to different Higgs boson particles. Three of them are neutral bosons, two scalars($h$ and $H$) and one pseudoscalar($A$), and the other two fields correspond to positively and negatively charged Higgs($H^{\pm}$).

Experimentally, Flavor-Changing Neutral currents(FCNC) are highly suppressed \cite{Gunion:1989we,Pedrini:2012vp}. However, these are allowed at tree-level in 2HDM. The model is thus constrained using the Paschos-Glashow–Weinberg condition, which states that the requirement of naturally flavor conserving currents implies that fermions of a given charge couple with a single scalar doublet only\cite{Glashow:1976nt, Paschos:1976ay}. A simple way to ensure such an arrangement is to impose discrete symmetries on the Lagrangian. One such Group is the $Z_2$ group. Under the imposition of this discrete symmetry on the Lagrangian, the scalar-field doublets transform as $\Phi_{1} \rightarrow \Phi_{1}$, $\Phi_{2} \rightarrow-\Phi_{2}$ or vice versa, and the quarks and leptons can have odd or even transformations over the same groups. Combination of different sets of these transformations lead to different variations in coupling of fermions with scalars. In general, these combinations can be classified into four major types of 2HDM, depending on how the fermions couple to the two scalar doublets. The Yukawa Lagrangian for these four major types of 2HDM can be expressed as

\begin{equation}
    \begin{aligned} 
\mathcal{L}_{\text {Yukawa }}^{2 \mathrm{HDM}}= & -\sum_{f=u, d, \ell} \frac{M_f}{v}\left(\kappa_h^f \bar{f} f h+\kappa_H^f \bar{f} f H-i \kappa_A^f \bar{f} \gamma_5 f A\right) \\ & -\left\{\frac{\sqrt{2} V_{u d}}{v} \bar{u}\left(m_u \kappa_A^u \mathrm{P}_L+m_d \kappa_A^d \mathrm{P}_R\right) d H^{+}+\frac{\sqrt{2} m_{\ell} \kappa_A^{\ell}}{v} \overline{\nu_L} \ell_R H^{+}+\text {H.c. }\right\},
\end{aligned}
\end{equation}
where the coupling constant values are mentioned in Table \ref{tab:couple}. These values can be understood in the context of the requirement to constrain the scalar potential

\begin{figure}[ht!]
    \centering
    \begin{subfigure}{0.4\linewidth}
        \centering
        \includegraphics[width=\textwidth]{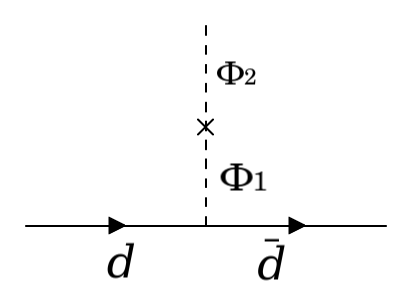}
        \caption{Soft $Z_2$ breaking term}
        \label{fig:soft}
    \end{subfigure}
    \begin{subfigure}{0.4\linewidth}
        \centering
        \includegraphics[width=\textwidth]{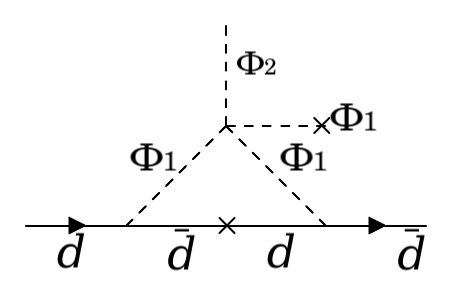}
        \caption{Hard $Z_2$ breaking term}
        \label{fig:hard}
    \end{subfigure}
    \caption{\small{Example of an effective Lagrangian terms which generate FCNC. These terms from soft $Z_2$ symmetry breaking \ref{fig:soft} have finite radiative corrections, whereas the ones from hard $Z_2$ symmetry breaking \ref{fig:hard} have infinite radiative corrections.\cite{Froggatt:1992wt}}}
    \label{fig:zbreak}
\end{figure}

In the strictest versions of these theories, the $Z_2$ symmetry applies to all terms in the potential. However, one can get away without any tree-level FCNC even if $Z_2$ is softly broken in the pure scalar sector\cite{Ginzburg:2004vp,sarkar}, which just yield finite corrections to masses. On the other hand, there can be quartic terms that break the $Z_2$ symmetry in the 'hard' way. These induce logarithmically divergent loop contributions (see Figure \ref{fig:zbreak}) to, say, $\Bar{d}d\Phi_2$ coupling in a scenario where $\Bar{d}d\Phi_1$ terms are set down at the tree level. This results in the appearance of a divergent part in the $\Bar{d}d\Phi_1$ coupling, which effectively violates natural flavor conservation at higher scales, via renormalization group(RG) evolution. The FCNC processes thus induced at different scales can take a toll on the 2HDM phenomenology\cite{Froggatt:1992wt,10.1143/PTP.63.234}. Additionally, the RG analysis of 2HDM also shows that soft $Z_2$ symmetry breaking extends the valid parameter space in the model, whereas an exact symmetry makes the parameter space of the model too constrained\cite{Oredsson:2018yho,Chakrabarty:2014aya}. The calculations and analyses here will thus be on the CP-conserving 2HDM with softly-breaking $Z_{2}$ symmetry.

Thus, for a CP-conserving 2HDM with softly breaking $Z_2$ symmetry, the general potential is 
\begin{equation}
    \begin{aligned}
V&=m_{11}^{2} \Phi_{1}^{\dagger} \Phi_{1}+m_{22}^{2} \Phi_{2}^{\dagger} \Phi_{2}-m_{12}^{2}\left(\Phi_{1}^{\dagger} \Phi_{2}+\Phi_{2}^{\dagger} \Phi_{1}\right) +\frac{\lambda_{1}}{2}\left(\Phi_{1}^{\dagger} \Phi_{1}\right)^{2}+\frac{\lambda_{2}}{2}\left(\Phi_{2}^{\dagger} \Phi_{2}\right)^{2}+\lambda_{3} \Phi_{1}^{\dagger} \Phi_{1} \Phi_{2}^{\dagger} \Phi_{2} \\
&+\lambda_{4} \Phi_{1}^{\dagger} \Phi_{2} \Phi_{2}^{\dagger} \Phi_{1}+\frac{\lambda_{5}}{2}\left[\left(\Phi_{1}^{\dagger} \Phi_{2}\right)^{2}+\left(\Phi_{2}^{\dagger} \Phi_{1}\right)^{2}\right]. \\
\end{aligned}
\end{equation}
When expressed in the terms of mass eigenstates of the charged and neutral Higgs bosons, the parameters $m_{11}^2, m_{22}^2$, and $\lambda_{1}, \lambda_{2}, \lambda_{3}, \lambda_{4}$ in the quartic potential can be written in terms of $\alpha$: the neutral CP-even Higgs diagonalizing mixing angle, tan$\beta=\frac{v_2}{v_1}$: the ratio of the two vev, and the masses $m_h$, $m_H$, $m_A$, $m_{H\pm}$ of all the Higgs Bosons. The mass $m_h$ is taken to be the 125 GeV Higgs boson. The Higgs decay width values will thus depend on these parameters, and have important phenomenological implications. The coupling constants of the three neutral Higgs bosons with the different fermions and bosons for the four types of 2HDM in the mass basis, as given in Table \ref{tab:couple} below.
\begin{table}[h!]
    \centering
    \begin{tabular}{|c|c|c|c|c|}
    \hline
         &  Type I & Type II & Type Lepton-specific(X) & Type Flipped(Y)\\
         \hline
      $\kappa^u_h$   & $\frac{\cos{\alpha}}{\sin{\beta}}$ & $\frac{\cos{\alpha}}{\sin{\beta}}$ & $\frac{\cos{\alpha}}{\sin{\beta}}$ & $\frac{\cos{\alpha}}{\sin{\beta}}$\\
      \hline
       $\kappa^d_h$   & $\frac{\cos{\alpha}}{\sin{\beta}}$ & $-\frac{\sin{\alpha}}{\cos{\beta}}$ & $\frac{\cos{\alpha}}{\sin{\beta}}$ & $-\frac{\sin{\alpha}}{\cos{\beta}}$\\
       \hline
       $\kappa^l_h$   & $\frac{\cos{\alpha}}{\sin{\beta}}$ & $-\frac{\sin{\alpha}}{\cos{\beta}}$ &  $-\frac{\sin{\alpha}}{\cos{\beta}}$ & $\frac{\cos{\alpha}}{\sin{\beta}}$\\
       \hline
        $\kappa^u_H$   & $\frac{\sin{\alpha}}{\sin{\beta}}$ & $\frac{\sin{\alpha}}{\sin{\beta}}$  & $\frac{\sin{\alpha}}{\sin{\beta}}$ & $\frac{\sin{\alpha}}{\sin{\beta}}$ \\
        \hline
        $\kappa^d_H$   & $\frac{\sin{\alpha}}{\sin{\beta}}$ & $\frac{\cos{\alpha}}{\cos{\beta}}$  & $\frac{\sin{\alpha}}{\sin{\beta}}$ & $\frac{\cos{\alpha}}{\cos{\beta}}$ \\
        \hline
        $\kappa^l_H$   & $\frac{\sin{\alpha}}{\sin{\beta}}$ & $\frac{\cos{\alpha}}{\cos{\beta}}$  & $\frac{\cos{\alpha}}{\cos{\beta}}$ & $\frac{\sin{\alpha}}{\sin{\beta}}$ \\
      \hline
      $\kappa^u_A$ & $\cot{\beta}$ & $\cot{\beta}$& $\cot{\beta}$ & $\cot{\beta}$\\
      \hline
      $\kappa^d_A$ & $-\cot{\beta}$ & $\tan{\beta}$& $-\cot{\beta}$ & $\tan{\beta}$\\
      \hline
      $\kappa^l_A$ & $-\cot{\beta}$ & $\tan{\beta}$& $\tan{\beta}$ & $-\cot{\beta}$\\
      \hline
      $\kappa^{W/Z}_h$ & \multicolumn{4}{|c|}{$\sin{(\beta-\alpha)}$}\\
      \hline
      $\kappa^{W/Z}_H$ & \multicolumn{4}{|c|}{$\cos{(\beta-\alpha)}$}\\
      \hline
      $\kappa^{H\pm}_h$ & \multicolumn{4}{|c|}{$\sin(\beta-\alpha)(m_h^2+2m_{H^\pm}^2-2\frac{m_{12}^2}{\sin\beta \cos\beta}) + 2\cos(\beta-\alpha)(m_h^2-\frac{m_{12}^2}{\sin\beta \cos\beta})\cot(2\beta)$}\\
      \hline
      $\kappa^{H\pm}_H$ & \multicolumn{4}{|c|}{$\cos(\beta-\alpha)(m_h^2+2m_{H^\pm}^2-2\frac{m_{12}^2}{\sin\beta \cos\beta}) + 2\sin(\beta-\alpha)(m_h^2-\frac{m_{12}^2}{\sin\beta \cos\beta})\cot(2\beta)$}\\
      \hline
    \end{tabular}
    \caption{Quark, lepton, vector boson and self Higgs couplings to neutral Higgs- h, H and A in the four types of 2HDM\cite{Branco:2011iw}. $\alpha$ is the mass matrix diagonalising mixing angle, and $\tan\beta$ is the ratio of the two vev of the scalar potential of 2HDM.}
    \label{tab:couple}
\end{table}
\section{The $\gamma\gamma$ and $Z\gamma$ decay modes in SM and 2HDM}
\subsection{Standard Model}
In the SM, the Higgs field doesn't interact with photons, which makes it massless. Thus, a Higgs boson can decay to photons only via loop induced diagrams where fermions as well as gauge bosons participate. 

The expression for Higgs to the $\gamma \gamma$ partial decay width at leading order is given by
\begin{equation}
    \Gamma(h \rightarrow \gamma \gamma)=\frac{G_F \alpha_{em}^2 M_H^3}{128 \sqrt{2} \pi^3}\left|\sum_f N_c Q_f^2 T_{1 / 2}^H\left(\tau_f\right)+T_1^H\left(\tau_W\right) \right|^2.
\end{equation}
where $G_F$ is the Fermi constant, $\alpha_{em}$ is the fine structure constant, $N_c$ is the number of colored charges, $Q_f$ is the charge of a particular fermion, $M_H, M_f, M_W$ are the Higgs mass, fermion masses and the W-boson mass, respectively. The summation over $f$ is for all fermion loops, and $\tau_f = \frac{M^2_H}{4M^2_f}$. The functions used are
\begin{equation}
    \begin{aligned}
T_{1 / 2}^H(x) & =2\left[\frac{1}{x}+\left(\frac{1}{x}-\frac{1}{x^2}\right) f(x)\right]\\
T_1^H(x) & =-\left[2+\frac{3}{x}+3(2 x-1) \frac{f(x)}{x^2}\right]\\
\end{aligned}
\end{equation}
where $f(x)$ is defined as
\begin{equation}
    f(x)= \begin{cases}\arctan ^2 \sqrt{\frac{x}{1-x}} & x \leq 1 \\ -0.25\left[\log \frac{1+\sqrt{1-x^{-1}}}{1-\sqrt{1-x^{-1}}}-i \pi\right]^2 & x>1\end{cases},\text{ and}
\end{equation}
The expression for Higgs to $Z\gamma$ partial decay width at leading order is given by
\begin{equation}
    \begin{aligned}
    \Gamma(h\rightarrow Z \gamma)=&\frac{G_F^2 M_W^2 \alpha_{em} M_H^3}{64 \pi^4}\left(1-\frac{M_Z^2}{M_H^2}\right)^3\left|\sum_f N_f \frac{Q_f \hat{v}_f}{c_W} T_{1 / 2}^H\left(\tau_f, \lambda_f\right)+T_1^H\left(\tau_W, \lambda_W\right)\right|^2
\end{aligned}
\end{equation}
where $\tau_i=4 M_i^2 / M_H^2 \text{ and } \lambda_i=4 M_i^2 / M_Z^2$, $M_Z$ being the mass of Z-boson, and the functions used are
\begin{equation}
    \begin{aligned}
T_{1 / 2}^H(x, y) & =\left[I_1(x, y)-I_2(x, y)\right] \\
T_1^H(x, y) & =\cos{\theta_W}\left\{4\left(3-\tan^2{\theta_W}\right) I_2(x, y)+\left[\left(1+\frac{2}{x}\right) \tan^2{\theta_W}-\left(5+\frac{2}{x}\right)\right] I_1(x, y)\right\}
\end{aligned}
\end{equation}
with $\hat{v}_f=2 I_f^3-4 Q_f \sin^2\theta_W$. The functions $I_1$ and $I_2$ are,
\begin{equation}
   \begin{aligned}
& I_1(x, y)=\frac{x y}{2(x-y)}+\frac{x^2 y^2}{2(x-y)^2}\left[f\left(x^{-1}\right)-f\left(y^{-1}\right)\right]+\frac{x^2 y}{(x-y)^2}\left[g\left(x^{-1}\right)-g\left(y^{-1}\right)\right] \\
& I_2(x, y)=-\frac{x y}{2(x-y)}\left[f\left(x^{-1}\right)-f\left(y^{-1}\right)\right]
\end{aligned} 
\end{equation}
where the function $g(x)$ is
\begin{equation}
    g(x)= \begin{cases}\sqrt{x^{-1}-1} \sin^{-1} \sqrt{x} & x < 1 \\ \frac{\sqrt{1-x^{-1}}}{2}\left[\log \frac{1+\sqrt{1-x^{-1}}}{1-\sqrt{1-x^{-1}}}-i \pi\right] & x\geq1\end{cases}
\end{equation}
Note that the fermion loops interfere destructively with the W-boson loops because of their antisymmetric properties.
The dominant contribution comes from the W-boson loops followed by the top-quark fermion loop, because of its heavy mass. 
All constants and mass values needed to calculate these partial decay widths are experimentally known. The SM thus gives a unique value for these partial decay widths.
\subsection{Two Higgs doublet models}
In 2HDM, what may yield insights into details of the model parameters include the $\gamma\gamma$, $Z\gamma$ decay rates of not only $h$, but also of $H$ and $A$. While the fermion loops in general contribute to all the decays, the $W$-loops are present for $h$- and $H$-decays only. In addition, $H^\pm$-driven loops contribute to the loop-induced decay widths of $h$ and $H$, but not $A$. These features lend richness to the study of the $\gamma\gamma$ and $Z\gamma$ channels, as we shall see below. \\
The partial decay width of $h$ and $H$ to $\gamma\gamma$ in the leading order are given by
\begin{equation}
    \Gamma(h/H \rightarrow \gamma \gamma)=\frac{G_F \alpha_{em}^2 M_H^3}{128 \sqrt{2} \pi^3}\left|\sum_f N_c\kappa_{h/H}^f Q_f^2 T_{1 / 2}^H\left(\tau_f\right)+\kappa^W_{h/H}T_1^H\left(\tau_W\right) + \kappa_{h/H}^{H\pm}T_0^H(\tau_{H\pm})\right|^2
    \label{eq: gg}
\end{equation}
where
\begin{equation}
    \begin{aligned}
T_0^H(x) & =-\left[\frac{1}{x} - \frac{f(x)}{x^2}\right]
\end{aligned}
\end{equation}

Also, $\tau_i = \frac{M^2_H}{4M^2_i}$.\\
The partial decay width of CP-even scalar to Z$\gamma$ in the leading order is given by
\begin{equation}
    \begin{aligned}
    \Gamma(h/H \rightarrow Z \gamma)=&\frac{G_F^2 M_W^2 \alpha_{em} M_H^3}{64 \pi^4}\left(1-\frac{M_Z^2}{M_H^2}\right)^3\\
    &\times\left|\sum_f N_c \kappa_{h/H}^f\frac{Q_f \hat{v}_f}{c_W} T_{1 / 2}^H\left(\tau_f, \lambda_f\right)+\kappa_{h/H}^WT_1^H\left(\tau_W, \lambda_W\right)+\kappa_{h/H}^{H^\pm}T_0^H(\tau_{H\pm},\lambda_{H\pm})\right|^2
    \label{eq: zg}
\end{aligned}
\end{equation}
where
\begin{equation}
T_0^H(x,y) = \frac{\cos{2\theta_W}}{\sin{\theta_W}}I_1(x, y)\frac{M_W^2}{M_{H^\pm}^2}
\end{equation}
$\theta_W$ being the electroweak mixing angle. The expressions are very similar to the SM result, with an important difference in the coupling constants. From Table \ref{tab:couple}, it can be seen that the magnitude of the contribution of fermions and W-boson loops depend on the parameters $\alpha$ and $\beta$. The contribution from charged Higgs loop depend additionally on the parameter $m_{12}^2$ and the unknown mass of charged Higgs boson, $m_{H^\pm}$. 

For the CP-odd scalar $A$, only the fermion loops contribute to the in the leading order of the decay width, yielding
\begin{equation}
    \Gamma(A\rightarrow \gamma \gamma)=\frac{G_F \alpha_{em}^2 M_A^3}{128 \sqrt{2} \pi^3}\left|\sum_f N_c\kappa_{A}^f Q_f^2 T_{1 / 2}^A\left(\tau_f\right)\right|^2, \text{ and}
\end{equation}
\begin{equation}
    \Gamma(A\rightarrow Z \gamma)=\frac{G_F^2 M_W^2 \alpha_{em} M_H^3}{64 \pi^4}\left(1-\frac{M_Z^2}{M_H^2}\right)^3\left|\sum_f N_f \kappa_{A}^f\frac{Q_f \hat{v}_f}{c_W} I_2\left(\tau_f, \lambda_f\right)\right|^2
\end{equation}
\begin{equation}
   \text{where        } T_{1 / 2}^A(x) =\frac{2f(x)}{x}
\end{equation}
The only unknown parameters controlling the decay width of $A$ are thus $m_A$ and $\tan\beta$.
\section{Results for the SM-like $h$}
\begin{figure}
    \centering  
    \includegraphics[width=0.5\linewidth]{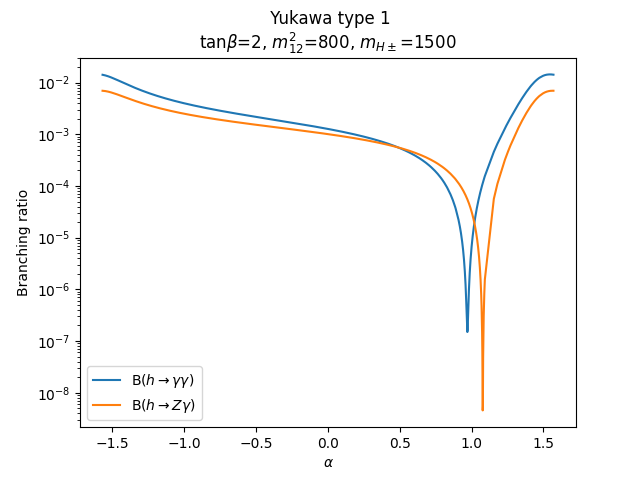}\hfill
    \includegraphics[width=0.5\linewidth]{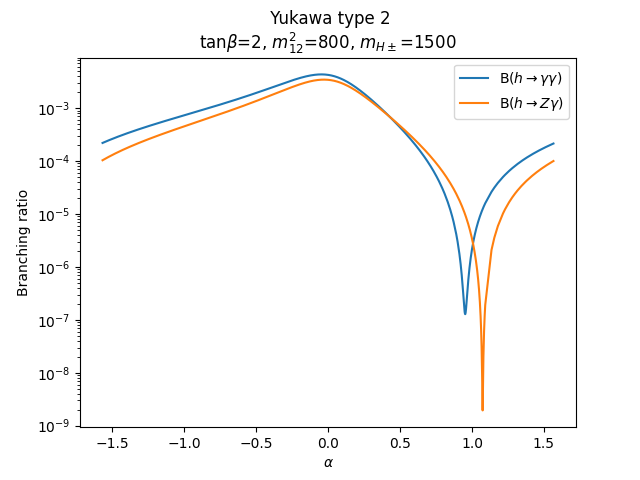}\hfill
    \includegraphics[width=0.5\linewidth]{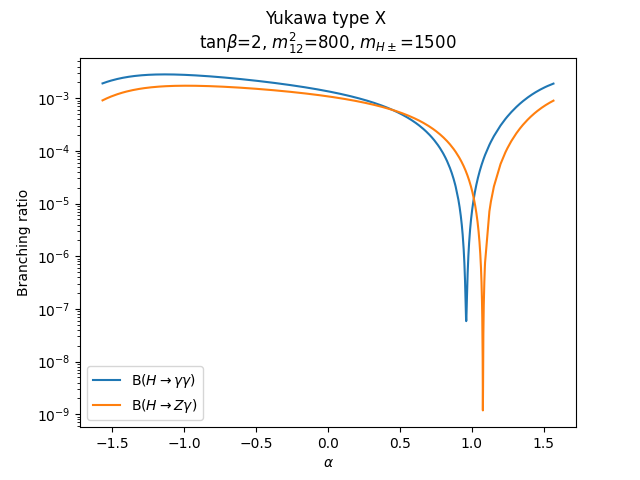}\hfill
    \includegraphics[width=0.5\linewidth]{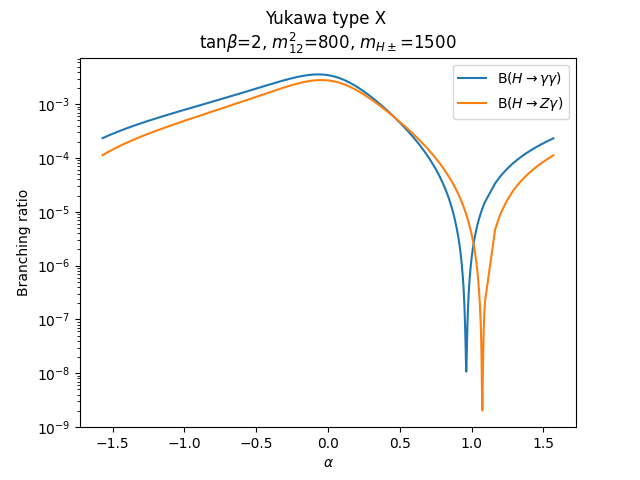}\hfill
    \caption{ $B(h\rightarrow Z\gamma)$ and $B(h\rightarrow \gamma\gamma)$ plotted against the mixing angle $\alpha$. Other parameter values are fixed with the following values: $\tan\beta$=2, $m_{12}^2$=800, $m_{H\pm}$=1500.}
    \label{fig:exp}
\end{figure}

\begin{figure}[H]
    \centering  
    \includegraphics[width=0.5\linewidth]{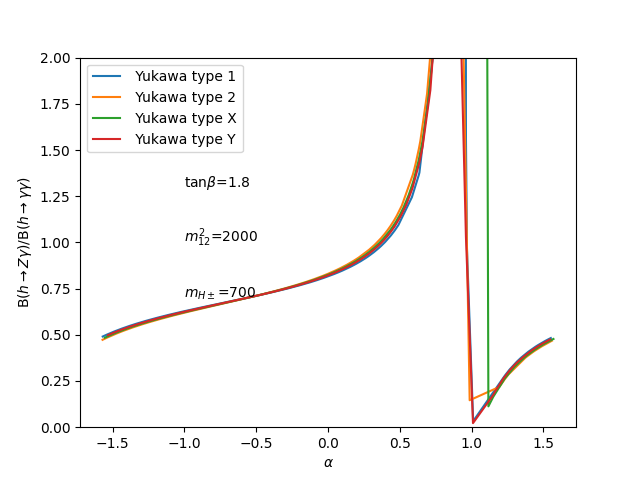}\hfill
    \includegraphics[width=0.5\linewidth]{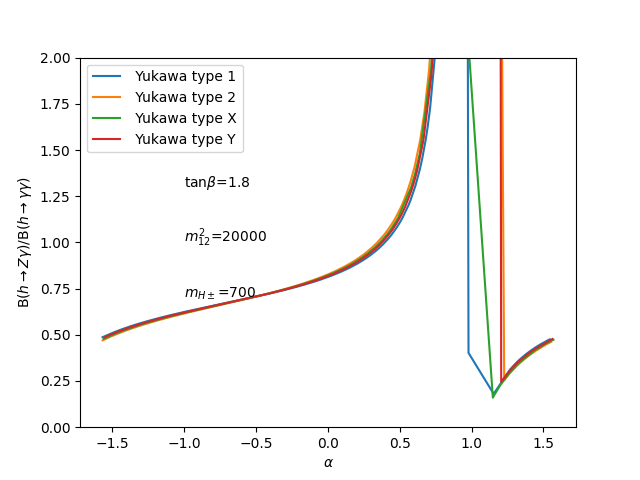}\hfill
    \includegraphics[width=0.5\linewidth]{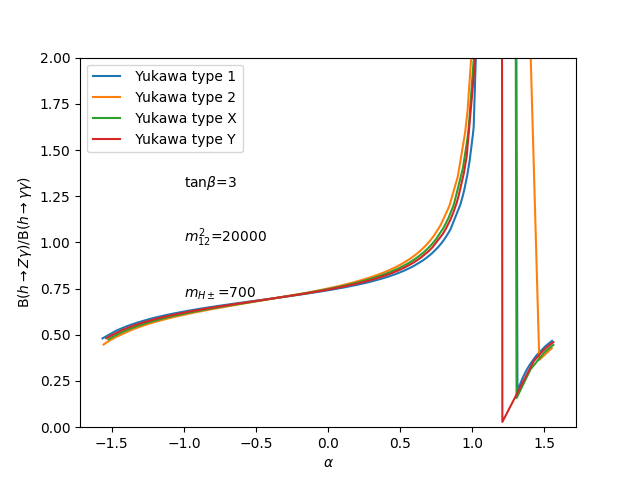}\hfill
    \includegraphics[width=0.5\linewidth]{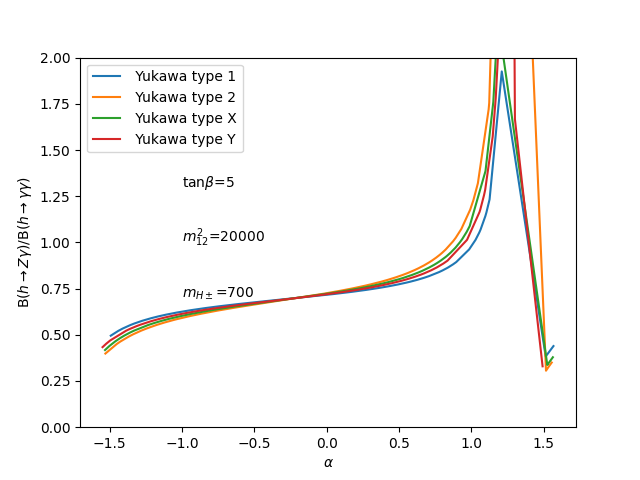}\hfill
    \includegraphics[width=0.5\linewidth]{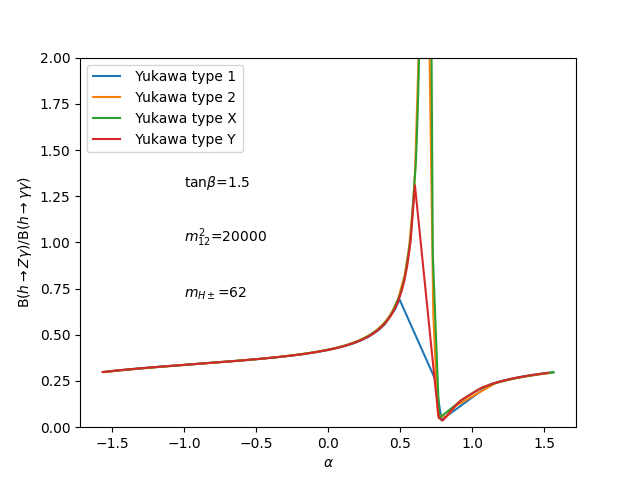}\hfill
    \includegraphics[width=0.5\linewidth]{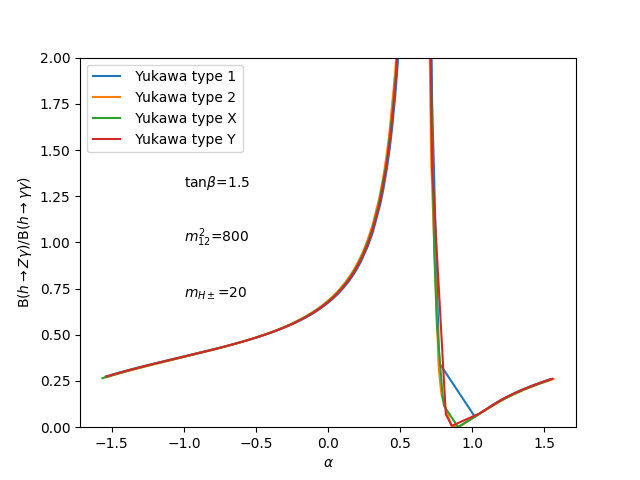}\hfill
    \includegraphics[width=0.5\linewidth]{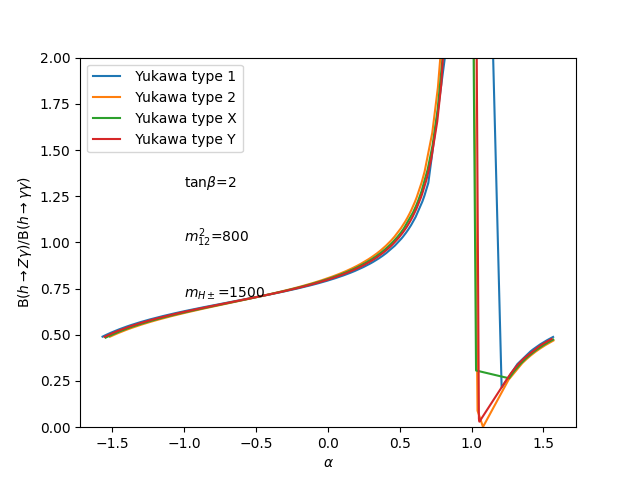}\hfill
    \includegraphics[width=0.5\linewidth]{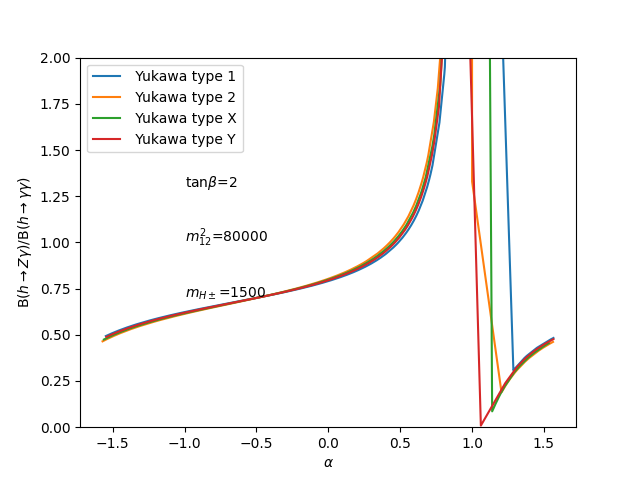}\hfill
    \caption{The ratio of the two branching ratios $B(h\rightarrow Z\gamma)$ and $B(h\rightarrow \gamma\gamma)$ plotted against the mixing angle $\alpha$, for different values of $\tan\beta, m_{12}^2, m_{H^\pm}$.}
    \label{fig:h}
\end{figure}

In the case of 125 GeV Higgs, the SM gives a unique prediction for the loop induced decays, namely, $B(h\rightarrow\gamma\gamma)>B(h\rightarrow Z\gamma)$. In 2HDM, however, with four unknown parameters- $\alpha, \beta, m_{12}^2, m_{H^\pm}$ for the light Higgs, this inequality reverses in some parameter space regions. This happens for parameters where the $W$-loop and $H^\pm$-loop contributors begin to get suppressed, but at a particular point $\Gamma(h\rightarrow\gamma\gamma)$ is subject to a stronger suppression than $\Gamma(h\rightarrow Z\gamma)$ and the latter thus becomes larger in value in comparison.\\
The above statement can be understood by considering figure \ref{fig:exp}, where the branching ratios of $\gamma\gamma$ and $Z\gamma$ decays are plotted against $\alpha$ for other fixed parameter values, $\tan\beta$=2, $m_{12}^2$=800, $m_{H\pm}$=1500. Both have a profile similar to the decay of $h\rightarrow WW$ as the dominant loop contributions are from W-bosons. $h$ couples to $W$ boson with a coupling strength proportional to $\sin(\beta-\alpha)$(Table \ref{tab:couple}), and $\Gamma(h\rightarrow WW)$ thus vanishes for $\beta=\alpha$. The $\gamma\gamma$ and $Z\gamma$ decays, however, have small yet significant contributions of fermion loops with different amplitudes as shown in Equation \ref{eq: gg} and \ref{eq: zg}. The two decay widths thus dip at different values of $\alpha$. This gives us a point in the parameter space where the dip for $\gamma\gamma$ precedes or succeeds the point of dip for $Z\gamma$, leading to the possibility of $B(h\rightarrow\gamma\gamma)<B(h\rightarrow Z\gamma)$ in the nearby region of values of $\alpha$ corresponding to the dip in $h\rightarrow \gamma\gamma$ rates (see Figure \ref{fig:exp}). Thus, there are regions in the parameter space where $B(h\rightarrow Z\gamma)>B(h\rightarrow \gamma\gamma)$ for all four types of 2HDM. (Figure \ref{fig:h})
However, such a prospect is disallowed over most of the parameter space by constraints on the scalar sector itself. These constraints are applied using \texttt{ScannerS}\cite{Muhlleitner:2020wwk}. The results are confirmed independently using \texttt{2HDMC}\cite{Eriksson:2009ws}, \texttt{Super-iso}\cite{Mahmoudi:2008tp}, \texttt{HiggsBounds}\cite{Bechtle:2008jh} and \texttt{HiggsSignals}\cite{Bechtle:2013xfa}. The sample regions of parameter space considered are the following:\\
$\alpha=[-\pi/2, \pi/2], \tan\beta = [0.1, 70], m_{12}^2 = [0.001,700000] GeV^2, m_{H^\pm} = [20 GeV, 2000 GeV]$.

\begin{figure}[H]
    \centering  
    \includegraphics[width=0.52\linewidth]{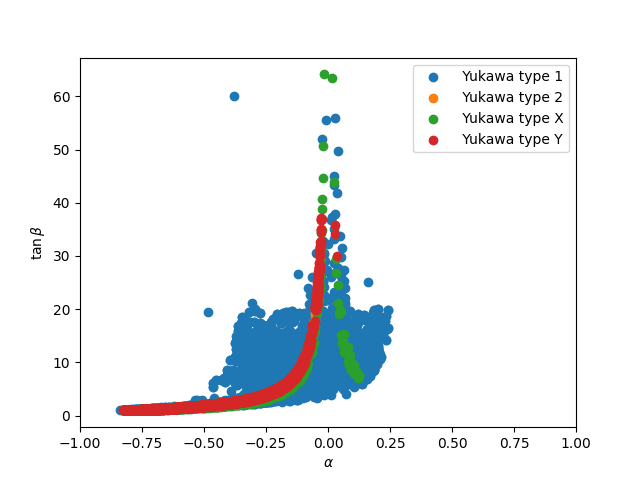}\hfill
    \caption{Sample regions of parameter space of the four different 2HDM types allowed by all theoretical and experimental constraints of 2HDM. $B(h\rightarrow \gamma\gamma)>B(h\rightarrow Z\gamma)$ everywhere.}
    \label{fig:h_al}
\end{figure}

\begin{figure}[H]
    \centering  
    \begin{subfigure}{0.47\linewidth}
        \centering
     \includegraphics[width=\textwidth]{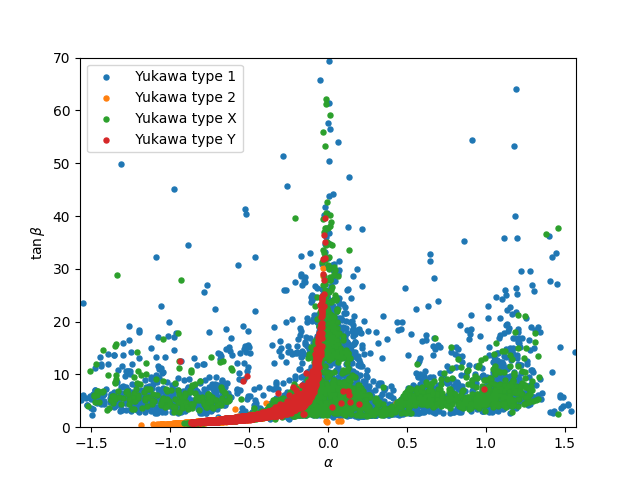}
     \caption{}
    \end{subfigure}  
    \begin{subfigure}{0.47\linewidth}
        \centering        \includegraphics[width=\textwidth]{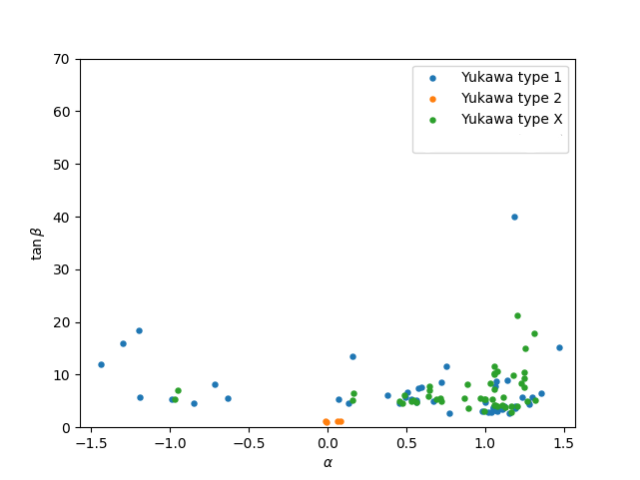}
        \caption{}
    \end{subfigure}
    \caption{(a)Sample regions of parameter space allowed by theoretical and experimental constraints of 2HDM, excluding the \texttt{HiggsSignals} data for 125 GeV scalar. (b)The subset of points  in “(a)” where $B(h1\rightarrow\gamma\gamma)<B(h\rightarrow Z\gamma)$, when data on $h$ are not used.}
    \label{fig:h_al_2}
\end{figure}

\begin{figure}
    \centering\includegraphics[width=0.5\linewidth]{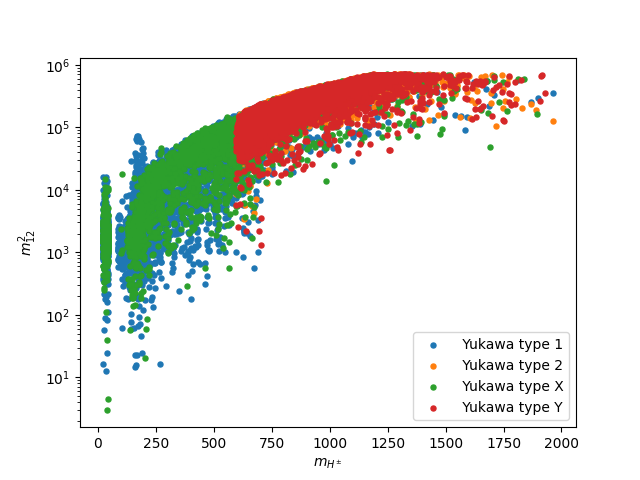}\hfill
    \includegraphics[width=0.5\linewidth]{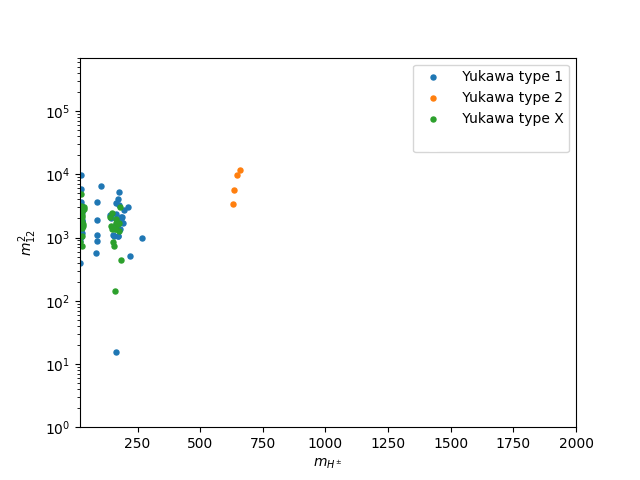}\hfill
    \includegraphics[width=0.5\linewidth]{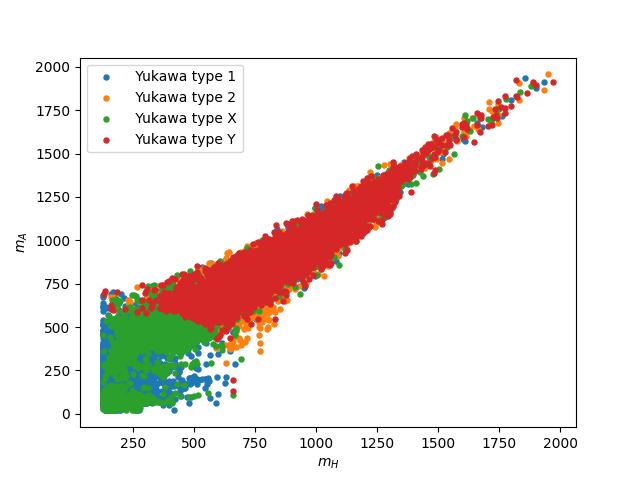}\hfill
    \includegraphics[width=0.5\linewidth]{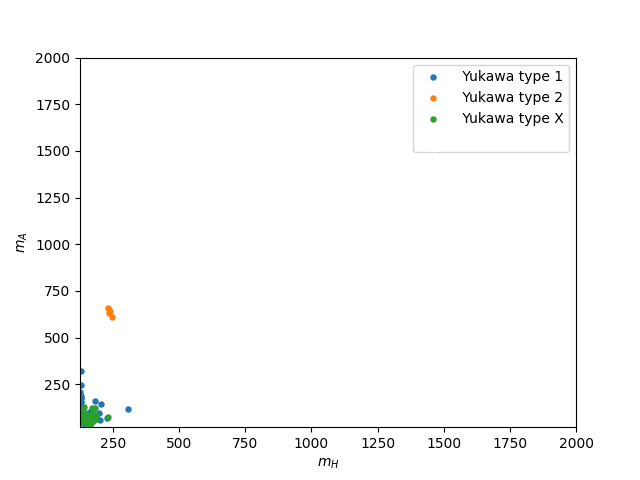}\hfill
    \caption{(Left) Other regions of the parameter space: $m_H, m_H\pm, m_A, m_{12}^2$, allowed by theoretical and experimental constraints of 2HDM, excluding the \texttt{HiggsSignals} data for 125 GeV scalar. (Right) The corresponding plots are subset points where $B(h\rightarrow Z\gamma)>B(h\rightarrow \gamma\gamma)$.}
    \label{fig:hnewc}
\end{figure}

When subjected to constraints of, Vacuum stability\cite{Espinosa:2013lma}, Perturbative Unitarity\cite{PhysRevD.16.1519}, Potential positivity, Electroweak precision constraints(EWPC)\cite{Peskin:1991sw}, flavor constraints\cite{doi:10.1146/annurev.nucl.012809.104534}, HiggsBounds and HiggsSignal measurement constraints, all regions of the parameter space where $B(h\rightarrow\gamma\gamma)<B(h\rightarrow Z\gamma)$ are excluded for all types of 2HDM (Figure \ref{fig:h_al}). This is mainly because the majority of the parameter space regions where $Z\gamma$ decay exceeds $\gamma \gamma$ are nearby points where the major decay channels of the 125 GeV scalar, $h\rightarrow WW$ and $h\rightarrow ZZ$, vanish or are highly suppressed. Experimental constraints thus exclude the parameter values.
However, even without considering the available data on the 125 GeV scalar, the theoretical and EWPC results, together with flavor constraints as well as analysis based on \texttt{HiggsBounds} ensure that the overwhelming majority of regions where $B(h\rightarrow\gamma\gamma)<B(h\rightarrow Z\gamma)$ are ruled out, with exception in a restricted parameter space regions for Type I, II and Type X 2HDM (Figure \ref{fig:h_al_2}). 
The theoretical and \texttt{HiggsBounds} constraints restrict $m_{H}$ and $m_{12}^2$ to an upper limit of $\sim$ 310 GeV and, $\sim$ 15000 $GeV^2$, respectively. The EWPC thus also restricts $m_A$ and $m_{H^\pm}$ to an upper limit (Figure \ref{fig:hnewc}). This is the reason exceptionally few points are allowed in Type II and none for Type-Y  as flavor constraints restrict $m_{H^\pm}\geq 600$ GeV for these two model types.\\
The above feature, namely most of the parameter spaces of 2HDM being disallowed by constraints excluding the 125-GeV scalar data, has been illustrated above for the state of brevity. However, it has been checked that a more exhaustive scan brings out the same feature.
\section{Results for $H$ and $A$}
\subsection{Decays of the CP-even heavy higgs $H$}
\begin{figure} 
    \includegraphics[width=0.5\linewidth]{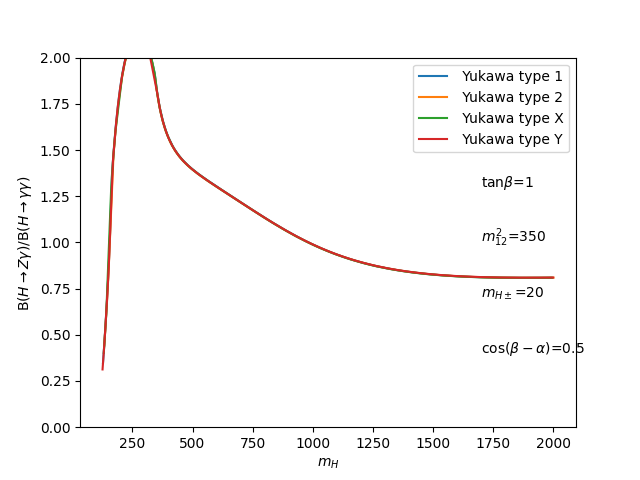}\hfill
    \includegraphics[width=0.5\linewidth]{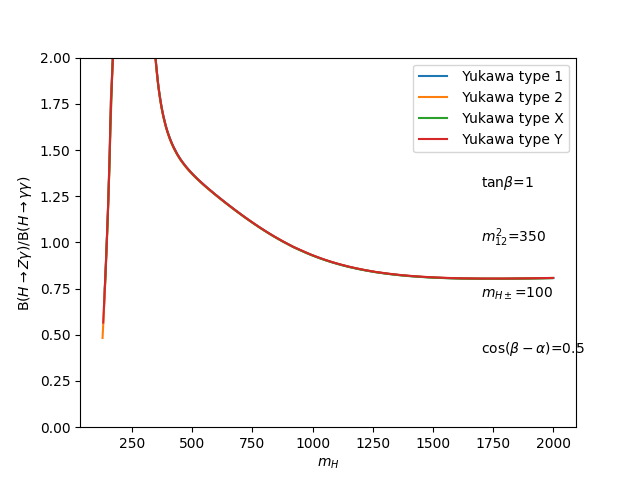}\hfill
    \includegraphics[width=0.5\linewidth]{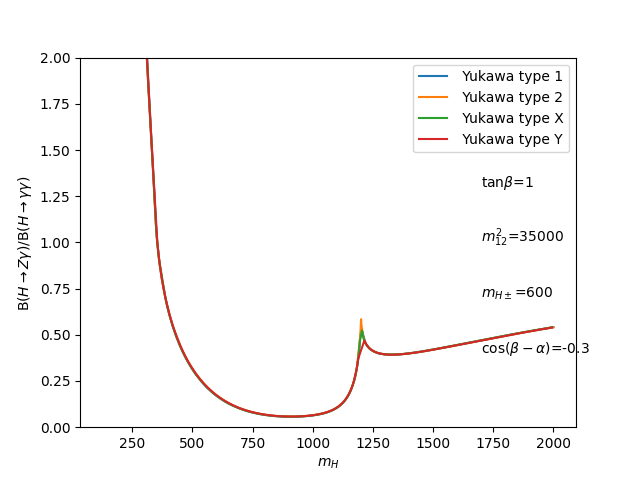}\hfill
    \includegraphics[width=0.5\linewidth]{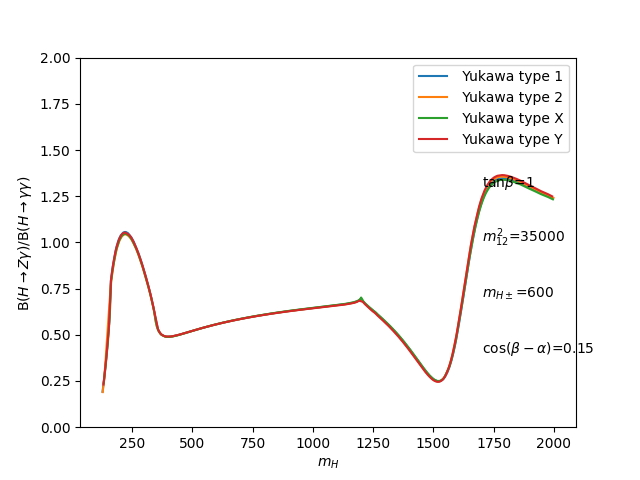}\hfill
    \includegraphics[width=0.5\linewidth]{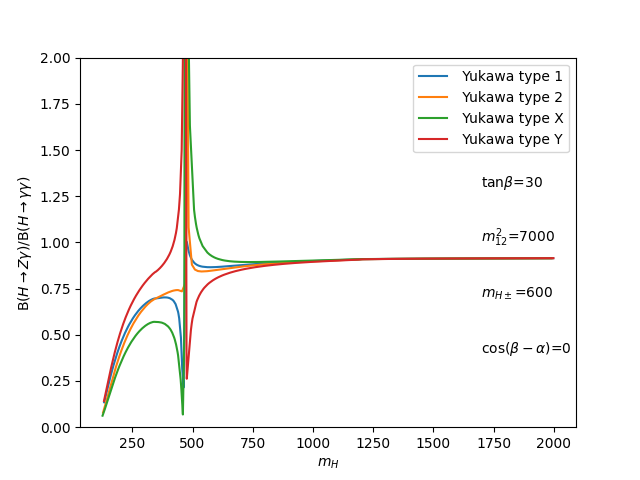}\hfill
    \includegraphics[width=0.5\linewidth]{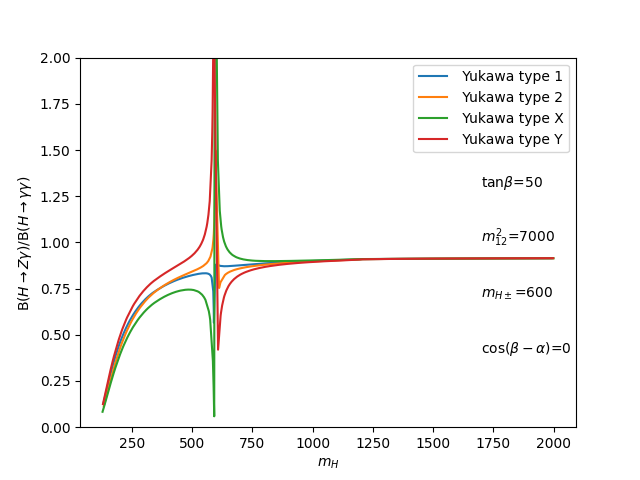}\hfill
    \includegraphics[width=0.5\linewidth]{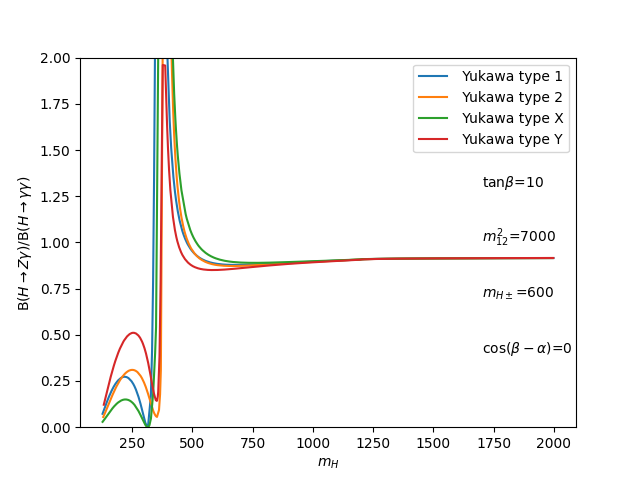}\hfill
    \includegraphics[width=0.5\linewidth]{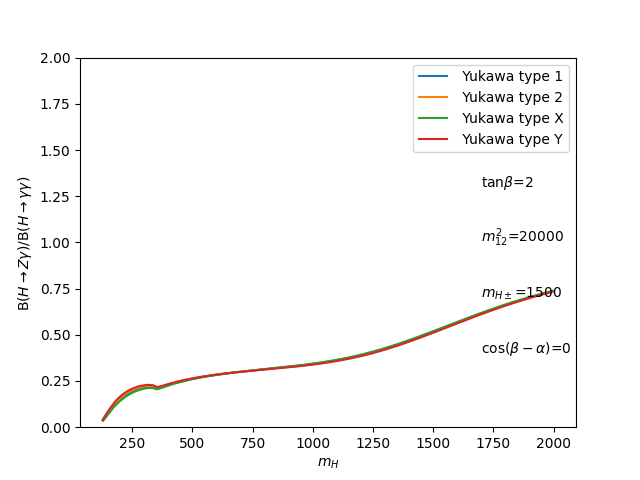}\hfill
    \caption{The ratio of the two branching ratios $B(H\rightarrow Z\gamma)$ and $B(H\rightarrow \gamma\gamma)$ plotted against $m_H$ for different values of $\tan\beta, m_{12}^2, m_{H^\pm}, \alpha$.}
    \label{HeavyB}
\end{figure}

\begin{figure}[H]
    \centering  
    \includegraphics[width=0.5\linewidth]{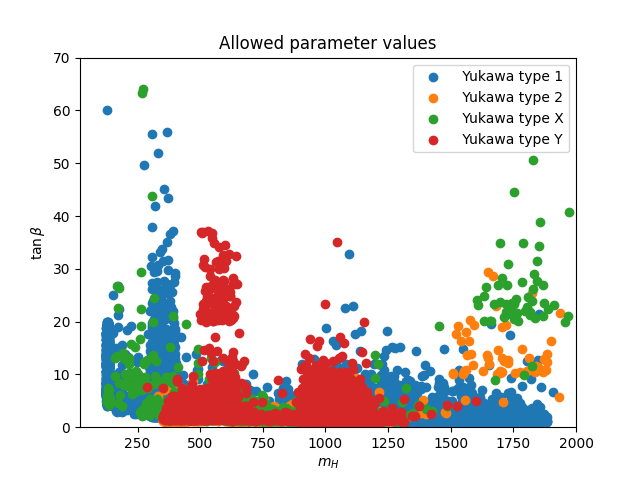}\hfill
    \includegraphics[width=0.5\linewidth]{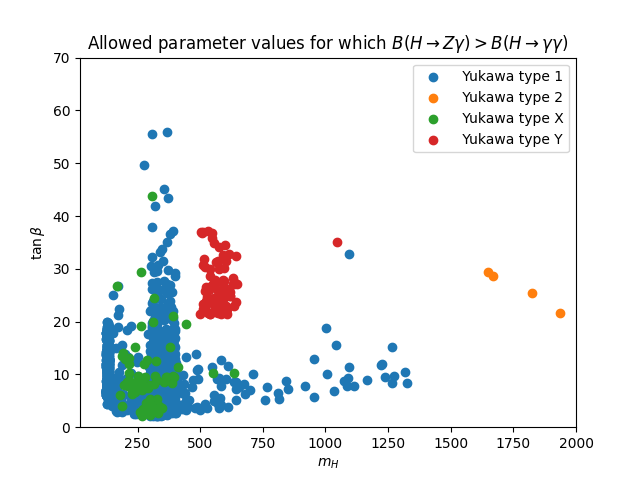}\hfill
    \includegraphics[width=0.5\linewidth]{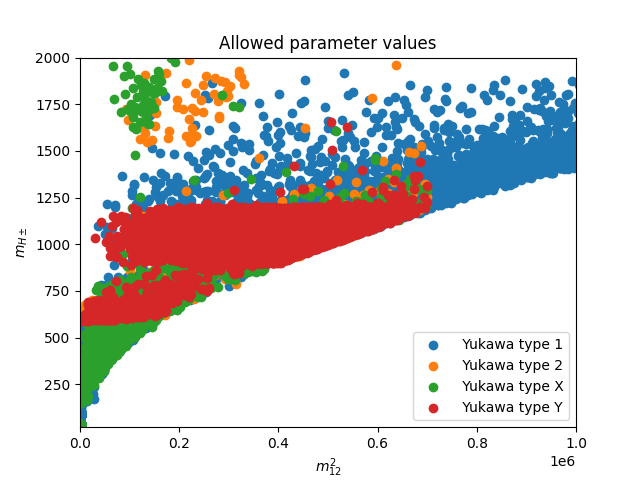}\hfill
    \includegraphics[width=0.5\linewidth]{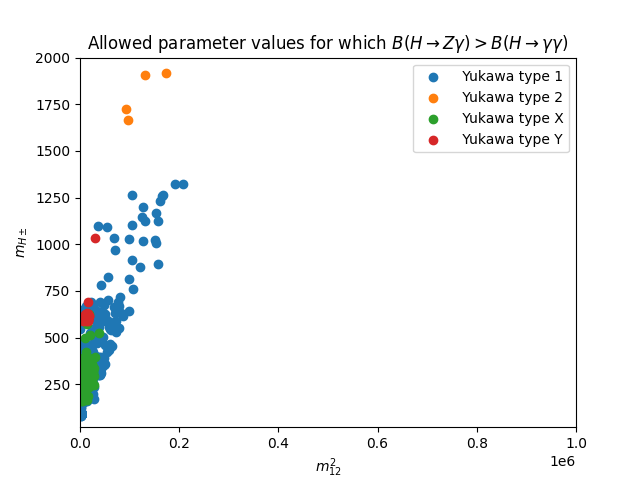}\hfill
    \includegraphics[width=0.5\linewidth]{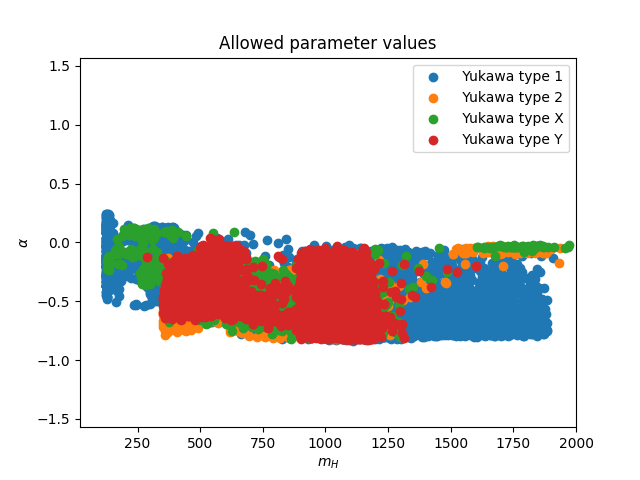}\hfill
    \includegraphics[width=0.5\linewidth]{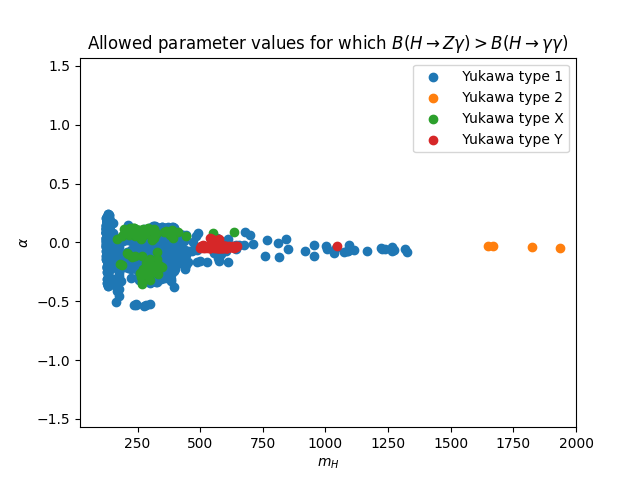}\hfill
    \includegraphics[width=0.5\linewidth]{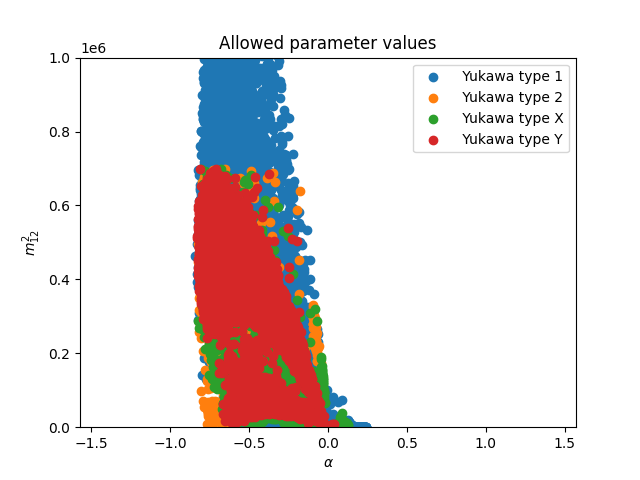}\hfill
    \includegraphics[width=0.5\linewidth]{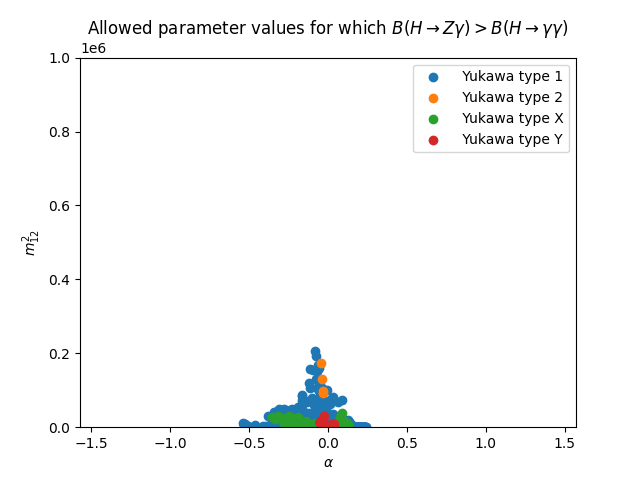}\hfill
    \caption{(Left) The regions of parameter space of $m_H, \tan\beta, m_{12}^2, m_{H^\pm}, \alpha$ allowed by theoretical and experimental constraints for all 4 types of 2HDM plotted against one another. The corresponding image on right are scatter points for which $\Gamma(H\rightarrow Z\gamma)>\Gamma(H\rightarrow\gamma\gamma)$.}
    \label{Hal}
\end{figure}

The branching ratios of the heavy scalar Higgs depend on five parameters of the model- $\alpha$, $\beta$, $m_{12}^2$, $m_{H^\pm}$, and $m_H$. The decay channels $H\rightarrow \gamma \gamma$ and $H\rightarrow Z\gamma$ become less significant with higher Higgs mass, as there are many other tree-level Higgs decays like $H\rightarrow tt$, $H\rightarrow hh$ with much higher partial decay widths that become significant. For $M_H\leq250 GeV$, the loop decays are relatively more significant in the regions of the parameter space, where the couplings to the gauge boson or heavy fermion pairs come with suppression factors arising out of parameters in the scalar potential.
\begin{figure}[H]
    \centering  
    \includegraphics[width=0.5\linewidth]{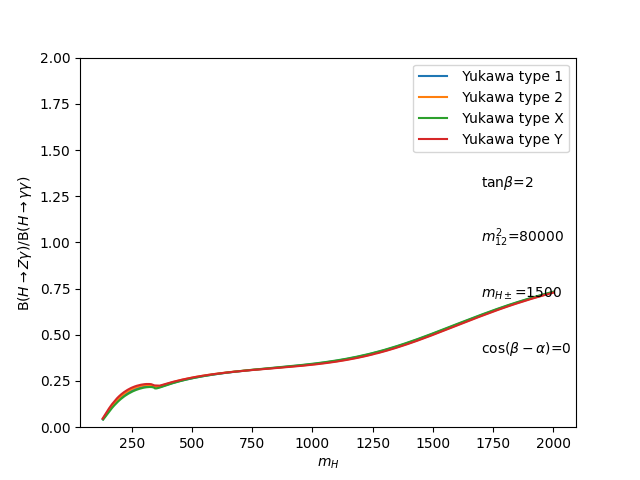}\hfill
    \includegraphics[width=0.5\linewidth]{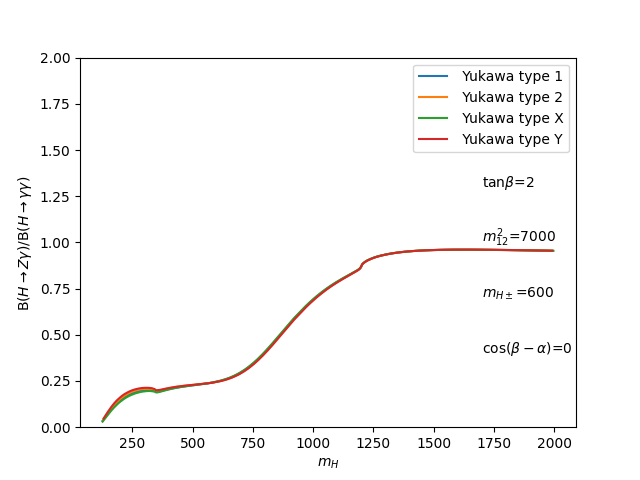}\hfill
    \includegraphics[width=0.5\linewidth]{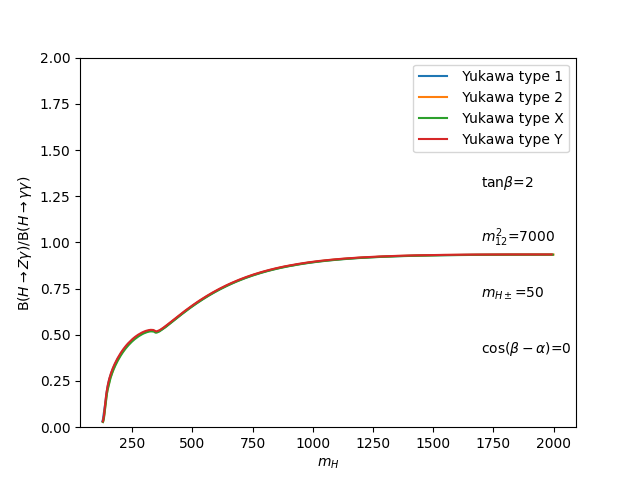}\hfill
    \includegraphics[width=0.5\linewidth]{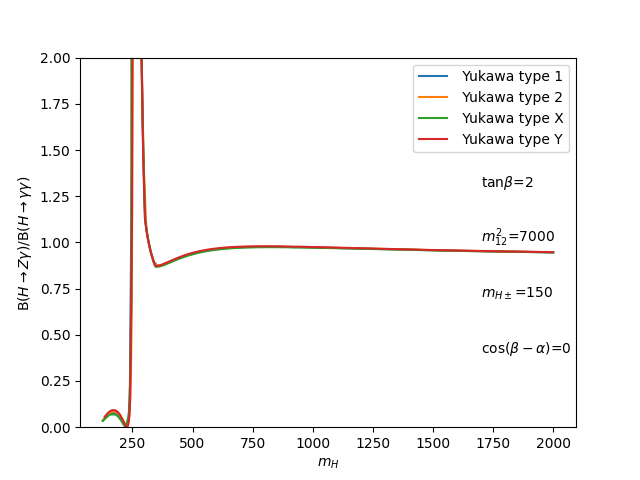}\hfill
    \caption{The ratio of the two branching ratios $B(H\rightarrow Z\gamma)$ and $B(H\rightarrow \gamma\gamma)$ plotted against $m_H$ for different values of $\tan\beta, m_{12}^2, m_{H^\pm}, \alpha$.}
    \label{HeavyA}
\end{figure}
For heavy scalar Higgs, there are possible regions in the parameter space where $B(H\rightarrow Z\gamma)>B(H\rightarrow \gamma\gamma)$ for all four types of 2HDM, as can be seen in Figure \ref{HeavyA} and \ref{HeavyB}. The five unknown parameters allow for varied possibilities in the values of decay amplitudes in equation \ref{eq: gg} and \ref{eq: zg}. \\
Note that in Figure \ref{HeavyA} and \ref{HeavyB}, the profile of $\frac{B(H\rightarrow Z\gamma)}{B(H\rightarrow \gamma\gamma)}$ are similar for all four types of 2HDM, particularly for large values of $m_H$. This is because the dominant contribution of the decay width comes from W-bosons loops, where the coupling strength is proportional to $\cos(\beta-\alpha)$(Table \ref{tab:couple}) in all model types. The contribution further increases when $m_H$ increases, thereby making the profile practically independent of the model type. The trilinear couplings with $H$ driving the charged Higgs loops, too, have the same parameter dependence in all four models. The slight differences for low values of $m_H$ is due to fermion loops having significant contribution to the decay amplitude, which vary for different model type.\\
Applying the theoretical and experimental constraints on the model, including \texttt{HiggsSignals}, allows $\Gamma(H\rightarrow Z\gamma)>\Gamma(H\rightarrow\gamma\gamma)$ for Higgs masses $m_H=[126 GeV, 2000 GeV]$ in all four types of 2HDM. This is showcased by plotting the five parameters against one another in Figure \ref{Hal}.
\subsection{CP-odd Higgs}
The CP-odd Higgs doesn't couple to the pair of vectors or charged scalar bosons. This makes its decay width such that major tree level and one-loop level decays depend only on two parameters of the 2HDM, namely, $\tan\beta$ and $m_A$. $A\rightarrow Zh$ and $A\rightarrow ZH$, if kinematically allowed, are the only major decay channels that depend on the parameter $\alpha$, in addition to $\beta$.
\begin{figure}[H]
    \centering  
    \includegraphics[width=0.5\linewidth]{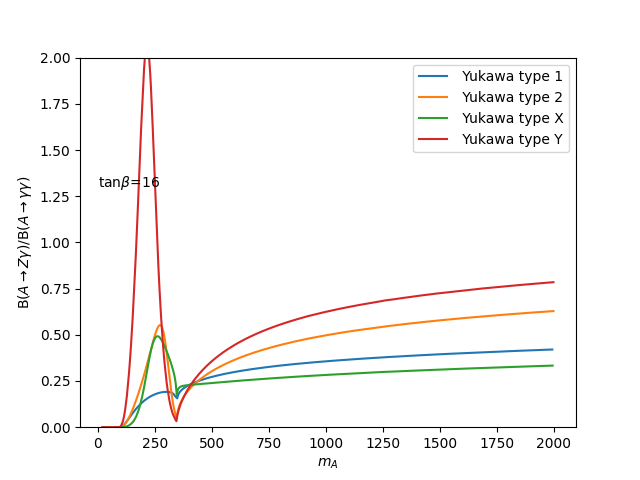}\hfill
    \includegraphics[width=0.5\linewidth]{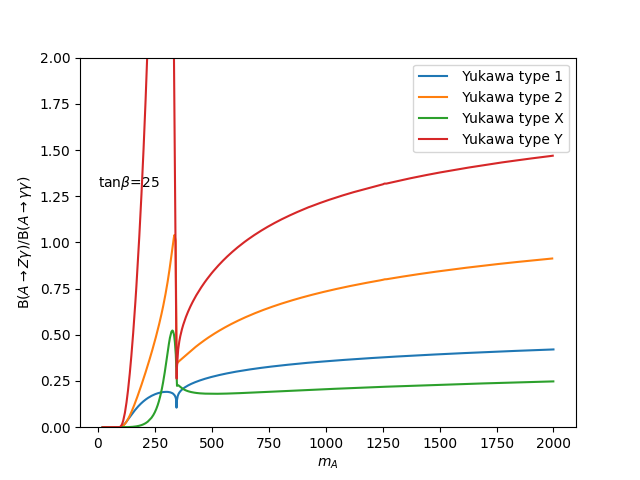}\hfill
    \includegraphics[width=0.5\linewidth]{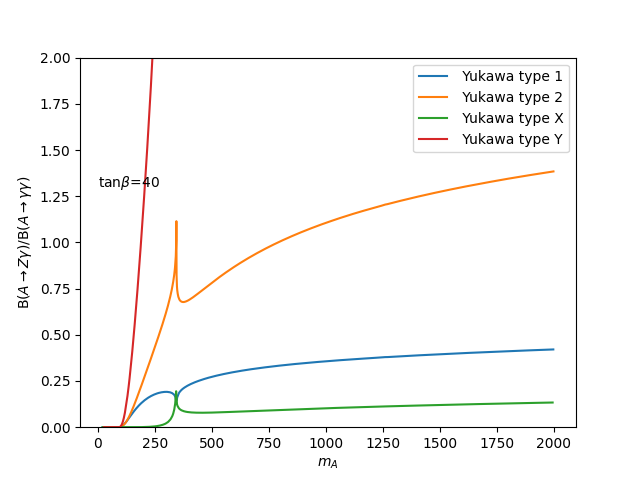}\hfill
    \includegraphics[width=0.5\linewidth]{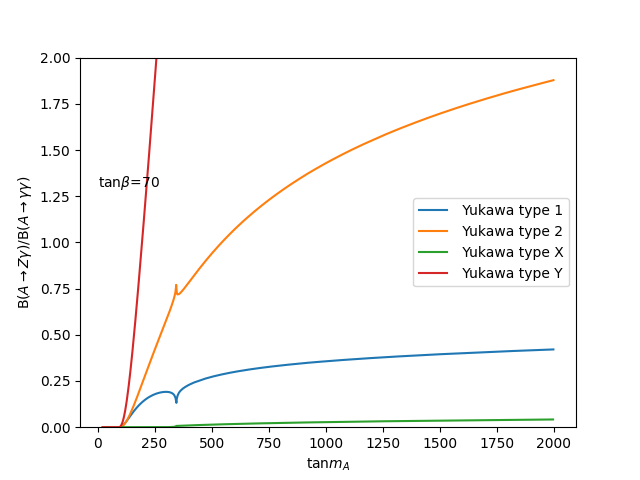}\hfill
    \caption{The ratio of the two branching ratios $B(A\rightarrow Z\gamma)$ and $B(A\rightarrow \gamma\gamma)$ plotted vs $m_A$ for different values of $\tan\beta$}
    \label{fig:A}
\end{figure}
The loop decays being mediated only by the fermions explains why at low Higgs mass these decays don't have high branching ratios at fermio-phobic regions.

\begin{figure}
    \centering  
    \includegraphics[width=0.5\linewidth]{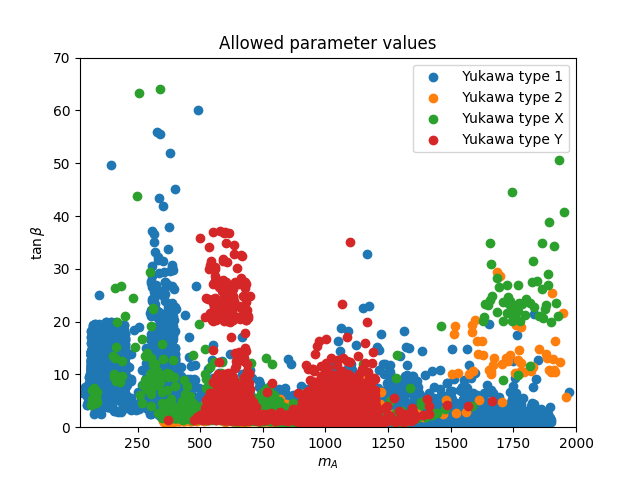}\hfill
    \includegraphics[width=0.5\linewidth]{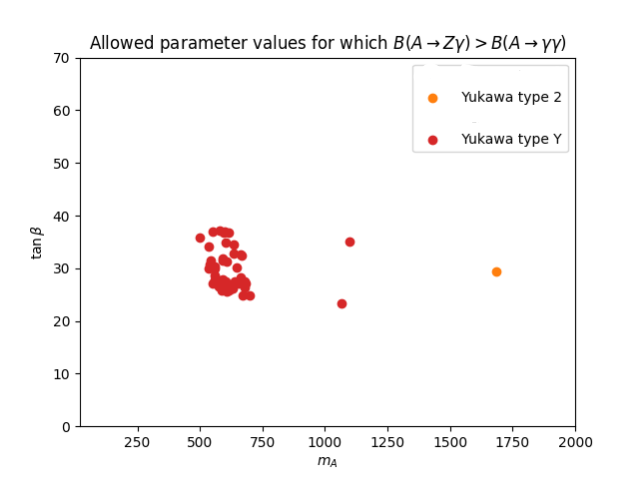}\hfill
    \caption{(Left) The regions of parameter space spanned by $\tan\beta$ and $m_A$ allowed by theoretical and experimental constraints for all four types of 2HDM. The remaining parameters in the scalar sector have been marginalized over. (Right) Points in the parameter space, in Types 1 and Y, where $\Gamma(A\rightarrow Z\gamma)>\Gamma(A\rightarrow\gamma\gamma)$.} 
    \label{fig:Aal}
\end{figure}

For the pseudoscalar Higgs, there are no regions in the parameter space where $B(A\rightarrow Z\gamma)>B(A\rightarrow \gamma\gamma)$ for Type I and Type-X 2HDM. This happens because the major fermionic contributions to the loop-induced decays are the heavier quarks. For all such quarks, the couplings to $A$ are proportional to $\cot\beta$ in Type I and X (see Table \ref{tab:couple}). Thus, there is no scope for some particular values of $\beta$ enhancing the contribution to $\Gamma(A\rightarrow Z\gamma)$ to values greater than those of $\Gamma(A\rightarrow\gamma\gamma)$. In Types II and Y, on the other hand, the coupling to the top quark goes as $\cot\beta$ whereas that to the bottom quark is proportional to $\tan\beta$. This makes the relative strengths of the contributions to both the decay widths variable, depending on $\beta$. It is found that some regions thus allow bigger contributions to $\Gamma(A\rightarrow Z\gamma)$ as compared to $\Gamma(A\rightarrow \gamma\gamma)$, for $\tan\beta > 20$, once all constraints are imposed (Figure \ref{fig:Aal}).\\
This leads us to an interesting conclusion: if a spin-0, CP-odd particle is identified in future experiments, which is found to have larger branching ratio to $Z\gamma$ than to $\gamma\gamma$, then one may identify the decaying particle as one belonging to Types II or Y, as opposed to Types I or X.
\section{Summary and Conclusions}
We have considered the four major kinds of 2HDM, namely, those of Type-I, Type-II, Type-X and Type-Y, and have studied the relative values of the $\gamma\gamma$ and $Z\gamma$ branching ratios of all the neutral scalar particles in the electroweak symmetry breaking sector of each one.  We have scanned the parameter space of each scenario over rather extensive ranges
of values of the scalar potential in each case. The scan has duly taken care of theoretical constraints such as the vacuum stability requirement as well as perturbative unitarity, electroweak precision constraints, and also experimental constraints from the Higgs data, searches for heavy scalars, and flavor constraints.

As far as the already discovered 125-GeV scalar is concerned, the diphoton channel events are already under scrutiny and are clearly more abundant than events in the $Z\gamma$ channel. Thus, any region of the parameter space in either model, which corresponds to, $\Gamma(h \rightarrow Z\gamma) \ge \Gamma(h \rightarrow \gamma\gamma)$ is automatically ruled out. Remarkably, the overwhelming majority of allowed regions  automatically ensure that the $\gamma\gamma$ branching ratio dominates over that of $Z\gamma$. This happens even without taking recourse to the analysis of data on the 125-GeV scalar; the \texttt{HiggsBounds} and flavor constraints do it together with Electroweak precision constraints. The limited regions of exceptions in Types I, II and X are eliminated by the 125-GeV data, using the code \texttt{HiggsSignals}.

As for the heavier scalar H, there are regions in all four types of models where the branching ratio in the $Z\gamma$ channel can exceed  that into $\gamma\gamma$.  This statement is true even after all constraints are taken into account.

The pseudoscalar A, however, always dominantly decays  into $\gamma\gamma$ in models of Type I and Type X. In Type II or Type Y, however, $Z\gamma$ rates may exceed those for diphotons after all constraints are applied, for $\tan\beta \ge 20$. Thus, for any discovered pseudoscalar which decays more frequently into $Z\gamma$ than to $\gamma\gamma$, the candidature
of Types I or Type X will have to be cancelled.
\printbibliography
\end{document}